\documentclass[11pt]{article} 
\usepackage[applemac]{inputenc}
\usepackage{cite}
\usepackage{amssymb}
\usepackage{amsmath}
\usepackage{graphicx}
\usepackage{relate}
\usepackage{fullpage}
\newcommand{\journal}[2]{#1}

\newcommand{\ceil}[1]{\left\lceil{#1}\right\rceil}

\newtheorem{lemma}{Lemma}
\newtheorem{theorem}{Theorem}

\newtheorem{proposition}{Proposition}

\newcommand{\qed}{\hfill\ensuremath{\Box}\medskip\\\noindent}
\newenvironment{proof}{\noindent\emph{Proof. }}{}

\newcommand{\depth}{\ensuremath{\mathrm{depth}}}

\newcommand{\parent}{\ensuremath{\mathrm{parent}}}

\newcommand{\lab}{\ensuremath{\mathrm{label}}}

\newcommand{\phrase}{\ensuremath{\mathrm{phrase}}}

\newcommand{\occ}{\ensuremath{\mathit{occ}}}
\newcommand{\reference}{\ensuremath{\mathrm{reference}}}

\newcommand{\rpre}{\ensuremath{\mathrm{rpre}}}
\newcommand{\rsuf}{\ensuremath{\mathrm{rsuf}}}
\newcommand{\lastmatch}{\ensuremath{\mathrm{lastmatch}}}
\newcommand{\zlw}{\rm \texttt{ZLW}}
\newcommand{\zla}{\rm \texttt{ZL78}}
\newcommand{\zlb}{\rm \texttt{ZL77}}
\newcommand{\deltab}{\ensuremath{\bar{\delta}}}

\title{Improved Approximate String Matching and Regular Expression Matching on Ziv-Lempel Compressed Texts\footnote{An extended abstract of this paper appeared in Proceedings of the 18th Annual Symposium on Combinatorial Pattern Matching, 2007.}}

\author{Philip Bille\thanks{IT University of Copenhagen, Rued Langgaards Vej 7, 2300
Copenhagen S, Denmark. Email: {\tt beetle@itu.dk}.} \and Rolf Fagerberg\thanks{University of Southern Denmark, Campusvej 55, 5230 Odense M, Denmark, Email: \texttt{rolf@imada.sdu.dk}.} \and Inge Li G{\o}rtz\thanks{Technical University of Denmark, Building 322, 2800 Kgs. Lyngby, Denmark. Email: {\tt ilg@imm.dtu.dk}.}}
\date{\today}

\begin{document}
\maketitle

\begin{abstract}
We study the approximate string matching and regular expression matching problem for the case when the text to be searched is compressed with the Ziv-Lempel adaptive dictionary compression schemes. We present a time-space trade-off that leads to algorithms improving the previously known complexities for both problems. In particular, we significantly improve the space bounds, which in  practical applications are likely to be a bottleneck.
\end{abstract}

\section{Introduction} 
Modern text databases, e.g. for biological and World Wide Web data, are huge. To save time and space, it is desireable if  data can be kept in compressed form and still allow efficient searching. Motivated by this Amir and Benson~\cite{AB1992a, AB1992} initiated the study of \emph{compressed pattern matching problems}, that is, given a text string $Q$ in compressed form $Z$ and a specified (uncompressed) pattern $P$, find all occurrences of $P$ in $Q$ without decompressing $Z$. The goal is to search more efficiently than the na{\"i}ve approach of decompressing $Z$ into $Q$ and then searching for $P$ in $Q$. Various compressed pattern matching algorithms have been proposed depending on the type of pattern and compression method, see e.g.,~\cite{AB1992, FT1998, KTSMA1998, KNU2003, Navarro2003, MUN2003}. For instance, given a string $Q$ of length $u$ compressed with the Ziv-Lempel-Welch scheme~\cite{Welch1984} into a string of length $n$, Amir et al.~\cite{ABF1996} gave an algorithm for finding all exact occurrences of a pattern string of length $m$ in $O(n + m^2)$ time and space. 

In this paper we study the classical approximate string matching and regular expression matching problems in the context of compressed texts. As in previous work on these problems~\cite{KNU2003, Navarro2003} we focus on the popular \zla\  and \zlw\  adaptive dictionary compression schemes~\cite{ZL1978, Welch1984}. We present a new technique that gives a general time-space trade-off. The resulting algorithms improve all previously known complexities for both problems. In particular, we significantly improve the space bounds. When searching large text databases, space is likely to be a bottleneck and therefore this is of crucial importance.

\subsection{Approximate String Matching}
Given strings $P$ and $Q$ and an
\emph{error threshold} $k$, the classical \emph{approximate string matching problem} is to find all ending positions of substrings of $Q$ whose \emph{edit distance} to $P$ is at most $k$. The edit distance between two strings is the minimum number of insertions, deletions, and substitutions needed to convert one string to the other. The classical dynamic programming solution due to Sellers~\cite{Sellers1980} solves the problem in $O(um)$ time and $O(m)$ space, where $u$ and $m$ are the length of $Q$ and $P$, respectively. Several improvements of this result are known, see e.g., the survey by Navarro~\cite{Navarro2001a}. For this paper we are particularly interested in the fast solution for small values of $k$, namely, the $O(uk)$ time algorithm by Landau and Vishkin~\cite{LV1989} and the more recent $O(uk^4/m + u)$ time algorithm due to Cole and Hariharan~\cite{CH2002} (we assume w.l.o.g. that $k < m$). Both of these can be implemented in $O(m)$ space.

Recently, K{\"a}rkk{\"a}inen et al.~\cite{KNU2003} studied this problem for text compressed with the \zla/\zlw\  compression schemes. If $n$ is the length of the compressed text, their algorithm achieves $O(nmk + \occ)$ time and $O(nmk)$ space, where $\occ$ is the number of occurrences of the pattern. Currently, this is the only non-trivial worst-case bound for the general problem on compressed texts. For special cases and restricted versions, other algorithms have been proposed~\cite{MKTSA2000, NR1998}.  An experimental study of
the problem and an optimized practical implementation can be found in~\cite{NKTSA01}.

In this paper, we show that the problem is closely connected to the uncompressed problem and we achieve a simple time-space trade-off. More precisely, let $t(m,u, k)$ and $s(m,u,k)$ denote the time and space, respectively, needed by any algorithm to solve the (uncompressed) approximate string matching problem with error threshold $k$ for pattern and text of length $m$ and $u$, respectively. We show the following result. 
\begin{theorem}\label{thm:approx}
Let $Q$ be a string compressed using {\zla } into a string $Z$ of length $n$ and let $P$ be a pattern of length $m$. Given $Z$, $P$, and a parameter $\tau \geq 1$, we can find all approximate occurrences of $P$ in $Q$ with at most $k$ errors in $O(n(\tau + m + t(m, 2m+2k,k)) + \occ)$ expected time and $O(n/\tau + m + s(m,2m+2k,k) + \occ)$ space.
\end{theorem}
The expectation is due to hashing and can be removed at an additional $O(n)$ space cost. In this case the bound also hold for \zlw\  compressed strings. We assume that the algorithm for the uncompressed problem produces the matches in sorted order (as is the case for all algorithms that we are aware of). Otherwise, additional time for sorting must be included in the bounds. To compare Theorem~\ref{thm:approx} with the result of Karkkainen et al.~\cite{KNU2003}, plug in the Landau-Vishkin algorithm and set $\tau = mk$. This gives an algorithm using $O(nmk + \occ)$ time and $O(n/mk + m + \occ)$ space. This matches the best known time bound while improving the space by a factor $\Theta(m^2k^2)$. Alternatively, if we plug in the Cole-Hariharan algorithm and set $\tau = k^4 + m$ we get an algorithm using $O(nk^4 + nm + \occ)$ time and $O(n/(k^4 + m) + m + \occ)$ space. Whenever $k = O(m^{1/4})$ this is $O(nm + \occ)$ time and $O(n/m + m + \occ)$ space. 

To the best of our knowledge, all previous non-trivial compressed pattern matching algorithms for \zla/\zlw\  compressed text, with the exception of a very slow algorithm for exact string matching by Amir et al.~\cite{ABF1996}, use $\Omega(n)$ space. This is because the algorithms explicitly construct the dictionary trie of the compressed texts. Surprisingly, our results show that for the \zla\  compression schemes this is not needed to get an efficient algorithm. Conversely, if very little space is available our trade-off shows that it is still possible to solve the problem without decompressing the text.

\subsection{Regular Expression Matching}

Given a regular expression $R$ and a string $Q$, the \emph{regular expression matching problem} is to find all ending position of substrings in $Q$ that matches a string in the language denoted by $R$. The classic textbook solution to this problem due to Thompson~\cite{Thomp1968} solves the problem in $O(um)$ time and $O(m)$ space, where $u$ and $m$ are the length of $Q$ and $R$, respectively. Improvements based on the Four Russian Technique or word-level parallelism are given in~\cite{Myers1992, BFC2005, Bille06}. 

The only solution to the compressed problem is due to Navarro~\cite{Navarro2003}. His solution depends on word RAM techniques to encode small sets into memory words, thereby allowing constant time set operations. On a unit-cost RAM with $w$-bit words this technique can be used to improve an algorithm by at most a factor $O(w)$. For $w = O(\log u)$ a similar improvement is straightforward to obtain for our algorithm and we will therefore, for the sake of exposition, ignore this factor in the bounds presented below. With this simplification Navarro's algorithm uses $O(nm^2 + \occ \cdot m\log m)$ time and $O(nm^2)$ space, where $n$ is the length of the compressed string. In this paper we show the following time-space trade-off:

\begin{theorem}\label{thm:regularex}
Let $Q$ be a string compressed using {\zla } or {\zlw } into a string $Z$ of length $n$ and let $R$ be a regular expression of length $m$. Given $Z$, $R$, and a parameter $\tau \geq 1$, we can find all occurrences of substrings matching $R$ in $Q$ in $O(nm(m + \tau) + \occ\cdot m \log m)$ time and $O(nm^2/\tau + nm)$ space.
\end{theorem}
If we choose $\tau = m$ we obtain an algorithm using $O(nm^2 + \occ\cdot m \log m)$ time and $O(nm)$ space. This matches the best known time bound while improving the space by a factor $\Theta(m)$. With word-parallel techniques these bounds can be improved slightly. \journal{The full details are given in Section~\ref{sec:wordparallel}.}{}

\subsection{Techniques}
If pattern matching algorithms for \zla\ or \zlw\ compressed texts use $\Omega(n)$ working space they can explicitly store the dictionary trie for the compressed text and apply any linear space data structure to it. This has proven to be very useful for compressed pattern matching. However, as noted by Amir et al.~\cite{ABF1996}, $\Omega(n)$ working space may not be feasible for large texts and therefore more space-efficient algorithms are needed. Our main technical contribution is a simple $o(n)$ data structure for \zla\ compressed texts. The data structure gives a way to compactly represent a subset of the trie which combined with the compressed text enables algorithms to quickly access relevant parts of the trie. This provides a general approach to solve compressed pattern matching problems in $o(n)$ space, which combined with several other techniques leads to the above results. \journal{}{Due to lack of space we have left out the details for regular expression matching. They can be found in 
the full version of the paper~\cite{BFG07}.}

\section{The Ziv-Lempel Compression Schemes}\label{zlc}
Let $\Sigma$ be an \emph{alphabet} containing $\sigma = |\Sigma|$ \emph{characters}.  A \emph{string} $Q$ is a sequence of characters from $\Sigma$. The \emph{length} of $Q$ is $u = |Q|$ and the unique string of length $0$ is denoted $\epsilon$. The $i$th character of $Q$ is denoted $Q[i]$ and the substring beginning at position $i$ of length $j-i+1$ is denoted $Q[i,j]$. The Ziv-Lempel algorithm from 1978~\cite{ZL1978} provides a simple and natural way to represent strings, which we describe below. Define a \emph{\zla\ compressed string} (abbreviated \emph{compressed string} in the remainder of the paper) to be a string of the form 
$$
Z = z_1 \cdots z_n = (r_1, \alpha_1)(r_2, \alpha_2) \ldots (r_n, \alpha_n), 
$$
where $r_i \in \{0, \ldots, i-1\}$ and $\alpha_i \in \Sigma$. Each pair $z_i = (r_i, \alpha_i)$ is a \emph{compression element}, and $r_i$ and $\alpha_i$ are the \emph{reference} and \emph{label} of $z_i$, denoted by $\reference(z_i)$ and $\lab(z_i)$, respectively. Each compression element \emph{represents} a string, called a \emph{phrase}. The phrase for $z_i$, denoted $\phrase(z_i)$, is given by the following recursion.
\begin{equation*}
\phrase(z_i) = 
\begin{cases}
  \lab(z_i)    & \text{if $\reference(z_i) = 0$}, \\
  \phrase(\reference(z_i))\cdot \lab(z_i)   & \text{otherwise}.
\end{cases}
\end{equation*}
The $\cdot$ denotes concatenation of strings. The compressed string $Z$ \emph{represents} the concatenation of the phrases, i.e., the string $\phrase(z_1)\cdots \phrase(z_n)$.

Let $Q$ be a string of length $u$. In \zla, the compressed string representing $Q$ is obtained by greedily parsing $Q$ from left-to-right with the help of a dictionary $D$. For simplicity in the presentation we assume the existence of an initial compression element $z_0 = (0, \epsilon)$ where $\phrase(z_0) = \epsilon$. Initially, let $z_0 = (0, \epsilon)$ and let  $D = \{\epsilon\}$. After step $i$ we have computed a compressed string $z_0 z_1 \cdots z_i$ representing $Q[1, j]$ and $D = \{\phrase(z_0), \ldots, \phrase(z_i)\}$. We then find the longest prefix of $Q[j+1, u-1]$ that matches a string in $D$, say $\phrase(z_k)$, and let $\phrase(z_{i+1}) = \phrase(z_k) \cdot Q[j+1 + |\phrase(z_k)|]$. Set $D = D \cup \{\phrase(z_{i+1})\}$ and let $z_{i+1} = (k, Q[j + 1 + |\phrase(z_{i+1})|])$. The compressed string $z_0 z_1\ldots z_{i+1}$ now represents the string $Q[1,j + |\phrase(z_{i+1})|])$ and $D = \{\phrase(z_0), \ldots, \phrase(z_{i+1})\}$. We repeat this process until all of $Q$ has been read.

\begin{figure}[t] 
  \centering \includegraphics[scale=.6]{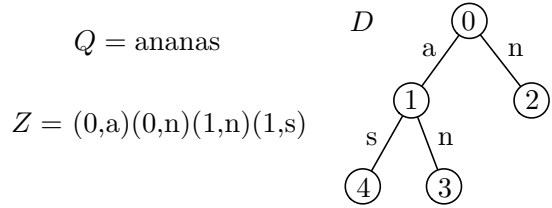}
  \caption{The compressed string $Z$ representing $Q$ and the corresponding dictionary trie $D$. Taken from~\cite{Navarro2003}.}
  \label{fig:lz78}
\end{figure}

Since each phrase is the concatenation of a previous phrase and a single character, the dictionary $D$ is prefix-closed, i.e., any prefix of a phrase is a also a phrase. Hence, we can represent it compactly as a trie where each node $i$ corresponds to a compression element $z_i$ and $\phrase(z_i)$ is the concatenation of the labels on the path from $z_i$ to node $i$. Due to greediness, the phrases are unique and therefore the number of nodes in $D$ for a compressed string $Z$ of length $n$ is $n+1$. An example of a string and the corresponding compressed string is given in Fig.~\ref{fig:lz78}.  

Throughout the paper we will identify compression elements with nodes in the trie $D$, and therefore we use standard tree terminology, briefly summed up here: The \emph{distance} between two elements is the number of edges on the unique simple path between them. The \emph{depth} of element $z$ is the distance from $z$ to $z_0$ (the root of the trie). An element $x$ is an \emph{ancestor} of an element $z$ if $\phrase(x)$ is a prefix of $\phrase(z)$. If also $|\phrase(x)| = |\phrase(z)| - 1$ then $x$ is the \emph{parent} of $z$.  If $x$ is ancestor of $z$ then $z$ is a \emph{descendant} of $x$ and if $x$ is the parent of $z$ then $z$ is the \emph{child} of $x$.The \emph{length} of a path $p$ is the number of edges on the path, and is denoted $|p|$. The \emph{label} of a path is the concatenation of the labels on these edges. 

Note that for a compression element $z$, $\reference(z)$ is a pointer to the parent of $z$ and $\lab(z)$ is the label of the edge to the parent of $z$. Thus, given $z$ we can use the compressed text $Z$ directly to decode the label of the path from $z$ towards the root in constant time per element. We will use this important property in many of our results.

If the dictionary $D$ is implemented as a trie it is straightforward to compress $Q$ or decompress $Z$ in $O(u)$ time. Furthermore, if we do not want to explicitly decompress $Z$ we can compute the trie in $O(n)$ time, and as mentioned above, this is done in almost all previous compressed pattern matching algorithm on Ziv-Lempel compression schemes. However, this requires at least $\Omega(n)$ space which is insufficient to achieve our bounds.  In the next section we show how to partially represent the trie in less space.

\subsection{Selecting Compression Elements}
Let $Z = z_0\ldots z_n$ be a compressed string. For our results we need an algorithm to select a compact subset of the compression elements such that the distance from any element to an element in the subset is no larger than a given threshold. More precisely, we show the following lemma.
\begin{lemma}\label{lem:special}
Let $Z$ be a compressed string of length $n$ and let $1 \leq \tau \leq n$ be parameter. There is a set of compression elements $C$ of $Z$, computable in $O(n\tau)$ expected time and $O(n/\tau)$ space with the following properties:
\begin{itemize}
  \item[(i)] $|C| = O(n/\tau)$.
  \item[(ii)] For any compression element $z_i$ in $Z$,  the minimum distance to any compression element in $C$ is at most $2\tau$. 
\end{itemize}
\end{lemma}
\begin{proof}
Let $1 \leq \tau \leq n$ be a given parameter. We build $C$ incrementally in a left-to-right scan of $Z$. The set is maintained as a dynamic dictionary using dynamic perfect hashing~\cite{DKMMRT1994}, i.e., constant time worst-case access and constant time amortized expected update. Initially, we set $C = \{z_0\}$. Suppose that we have read $z_0, \ldots, z_i$. To process $z_{i+1}$ we follow the path $p$ of references until we encounter an element $y$ such that $y \in C$.  We call $y$ the \emph{nearest special element} of $z_{i+1}$. Let $l$ be the number of elements in $p$ including $z_{i+1}$ and $y$. Since each lookup in $C$ takes constant time the time to find the nearest special element is $O(l)$. If $l < 2\cdot \tau$ we are done. Otherwise, if $l = 2\cdot \tau$, we find the $\tau$th element $y'$ in the reference path and set $C := C \cup \{y'\}$. As the trie grows under addition of leaves condition (ii) follows. Moreover, any element chosen to be in $C$ has at least $\tau$ descendants of distance at most $\tau$ that are not in $C$ and therefore condition (i) follows. The time for each step is $O(\tau)$ amortized expected and therefore the total time is $O(n\tau)$ expected. The space is proportional to the size of $C$ hence the result follows. \qed
\end{proof}

\subsection{Other Ziv-Lempel Compression Schemes}

A popular variant of \zla\  is the \zlw\ compression scheme~\cite{Welch1984}. Here, the label of compression elements are not explicitly encoded, but are defined to be the first character of the next phrase. Hence, \zlw\ does not offer an asymptotically better compression ratio over \zla\ but gives a better practical performance. The \zlw\ scheme is implemented in the UNIX program \texttt{compress}. From an algorithmic viewpoint \zlw\ is more difficult to handle in a space-efficient manner since labels are not explicitly stored with the compression elements as in \zla. However, if $\Omega(n)$ space is available then we can simply construct the dictionary trie. This gives constant time access to the label of a compression elements and therefore \zla\ and \zlw\ become "equivalent". This is the reason why Theorem~\ref{thm:approx} holds only for \zla\ when space is $o(n)$ but for both when the space is $\Omega(n)$. 

Another well-known variant is the \zlb\ compression scheme~\cite{ZL1977}. Unlike \zla\ and \zlw\ phrases in the \zlb\ scheme can be any substring of text that has already been processed. This makes searching much more difficult and none of the known techniques  for \zla\ and \zlw\ seems to be applicable. The only known algorithm for pattern matching on \zlb\ compressed text is due to Farach and Thorup~\cite{FT1998} who gave an algorithm for the exact string matching problem.

\section{Approximate String Matching}\label{approx}
In this section we consider the compressed approximate string matching problem. 
Before presenting our algorithm we need a few definitions and properties of approximate string matching. 

Let $A$ and $B$ be strings. Define the \emph{edit distance} between $A$ and $B$, $\gamma(A,B)$, to be the minimum number of insertions, deletions, and substitutions needed to transform $A$ to $B$. We say that $j \in [1, |S|]$ is a \emph{match with error at most $k$} of $A$ in a string $S$ if there is an $i \in [1, j]$ such that $\gamma(A, S[i,j]) \leq k$. Whenever $k$ is clear from the context we simply call $j$ a \emph{match}. All positions $i$ satisfying the above property are called a \emph{start} of the match $j$. The set of all matches of $A$ in $S$ is denoted $\Gamma(A, S)$. 
We need the following well-known property of approximate matches.
\begin{proposition}\label{prop:match}
Any match $j$ of $A$ in $S$ with at most $k$ errors must start in the interval $[\max(1, j-|A|+1-k), \min(|S|, j-|A|+1+k)]$.
\end{proposition}

\begin{proof}
Let $l$ be the length of a substring $B$ matching $A$ and ending at $j$. If the match starts outside the interval then either $l < |A| - k$ or $l > |A| + k$. In these cases, more than $k$ deletions or $k$ insertions, respectively, are needed to transform $B$ to $A$. \qed
\end{proof}

\subsection{Searching for Matches}

Let $P$ be a string of length $m$ and let $k$ be an error threshold. To avoid trivial cases we assume that $k < m$. 
Given a compressed string $Z = z_0z_1\ldots z_n$ representing a string $Q$ of length $u$ we show how to find $\Gamma(P, Q)$ efficiently. 

\begin{figure}[t] 
  \centering \includegraphics[scale=.5]{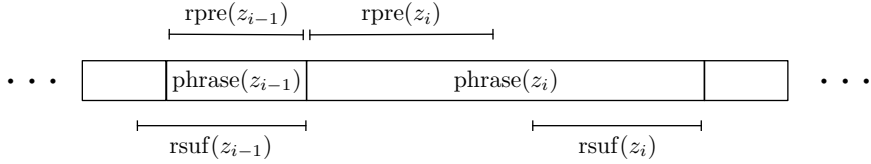}
  \caption{The relevant prefix and the relevant suffix of two phrases in $Q$. Here, $|\phrase(z_{i-1})| < m+k$ and therefore $\rsuf(z_{i-1})$ overlaps with previous phrases.}
  \label{fig:relevant}
\end{figure}

Let $l_i = |\phrase(z_i)|$, let $u_0 = 1$, and let $u_i = u_{i-1} + l_{i-1}$, for $1\leq i \leq n$, i.e., $l_i$ is the length of the $i$th phrase and $u_i$ is the starting position in $Q$ of the $i$th phrase. We process $Z$ from left-to-right and at the $i$th step we find all matches in $[u_i, u_i + l_i-1]$. Matches in this interval can be either \emph{internal} or \emph{overlapping} (or both). A match $j$ in $[u_i, u_i + l_i-1]$ is internal if it has a starting point in $[u_i, u_i + l_i-1]$ and overlapping if it has a starting point in $[1, u_i -1]$. To find all matches we will compute the following information for $z_i$.  
\begin{itemize}
  \item The start position, $u_i$, and length, $l_i$, of $\phrase(z_i)$.
  \item The \emph{relevant prefix}, $\rpre(z_i)$, and the \emph{relevant suffix}, $\rsuf(z_i)$, where 
  \begin{align*}
\rpre(z_i) &= Q[u_i, \min(u_i + m + k-1, u_i + l_i - 1)]\;,   \\ 
\rsuf(z_i) &= Q[\max(1, u_i +l_i - m - k ), u_i + l_i-1]\;.    
\end{align*}
In other words, $\rpre(z_i)$ is the largest prefix of length at most $m+k$ of $\phrase(z_i)$ and $\rsuf(z_i)$ is the substring of length $m+k$ ending at $u_i + l_i -1$. For an example see Fig.~\ref{fig:relevant}.
\item The \emph{match sets} $M_I(z_i)$ and $M_O(z_i)$, where 
\begin{align*}
    M_I(z_i) &= \Gamma(P, \phrase(z_i))\;, \\
    M_O(z_i) &= \Gamma(P, \rsuf(z_{i-1}) \cdot \rpre(z_i))\;. 
\end{align*}
We assume that both sets are represented as sorted lists in increasing order.
\end{itemize}

We call the above information the \emph{description} of $z_i$. In the next section we show how to efficiently compute descriptions. For now, assume that we are given the description of $z_i$. Then, the set of matches in $[u_i, u_i + l_i-1]$ is reported as the set
\begin{eqnarray*}
M(z_i) &=& \{j+u_i - 1 \mid j\in M_I(z_i)\}  \cup \\
&& \{j+u_i-1-|\rsuf(z_{i-1})| \mid j\in M_O(z_i) \cap [u_i, u_i + l_i-1]\}\;.
\end{eqnarray*}

We argue that this is the correct set. Since $\phrase(z_i) = Q[u_i, u_i + l_i-1]$ we have that $$j \in M_I(z_i) \Leftrightarrow   j + u_i - 1 \in \Gamma(P, Q[u_i, u_i + l_i-1]\;.$$ Hence, the set $\{j+u_i - 1\mid j\in M_I(z_i)\}$ is the set of all internal matches. Similarly, $\rsuf(z_{i-1}) \cdot \rpre(z_i) = Q[u_i - |\rsuf(z_{i-1})|, u_i + |\rpre(z_i)| - 1]$ and therefore 
$$j \in M_O(z_i) \Leftrightarrow   j + u_i - 1 - |\rsuf(z_{i-1})| \in \Gamma(P, Q[u_i - |\rsuf(z_{i-1})|, u_i + 1 + |\rpre(z_i)|])\;.$$
By Proposition~\ref{prop:match} any overlapping match must start at a position within the interval $[\max(1, u_i-m+1-k), u_i]$. Hence, $\{j + u_i -1- |\rsuf(z_{i-1})| \mid j\in M_O(z_i)\}$ includes all overlapping matches in $[u_i, u_i + l_i-1]$. Taking the intersection with $[u_i, u_i + l_i-1]$ and the union with the internal matches it follows that the 
set $M(z_i)$ is precisely the set of  matches in $[u_i, u_i + l_i-1]$. 
For an example see Fig.~\ref{fig:approx-description}. 

\begin{figure}[t]
\begin{center}
$$Q=\textrm{ananasbananer}, \quad P=\textrm{base}, \quad Z=\textrm{(0,a)(0,n)(1,n)(1,s)(0,b)(3,a)(2,e)(0,r)}$$
{\bf Descriptions}

$
\begin{array}{l  @{\hspace{5pt}} | @{\hspace{5pt}} l @{\hspace{5pt}} l @{\hspace{5pt}} l @{\hspace{5pt}} l @{\hspace{5pt}} l @{\hspace{5pt}} l @{\hspace{5pt}} l @{\hspace{5pt}} l}
 \hline & z_0 & z_1 & z_2 & z_3 & z_4 & z_5 & z_6 & z_7 \\ \hline
  u_i & 1 & 2& 3 & 5& 7& 8 & 11 & 13 \\
  l_i & 1 & 1 & 2 & 2 & 1& 3 & 2 & 1 \\
 \textrm{rpre}(z_i) & \textrm{a} & \textrm{n} & \textrm{an} & \textrm{as} & \textrm{b} & \textrm{ana} & \textrm{ne} & \textrm{r} \\
 \textrm{rsuf}(z_i) & \textrm{a} & \textrm{an} & \textrm{anas} & \textrm{ananas} & \textrm{nanasb} & \textrm{asbana} & \textrm{banane} & \textrm{ananer} \\
 M_I(z_i) & \emptyset & \emptyset & \emptyset &\{2\}& \emptyset & \emptyset & \emptyset & \emptyset \\
 M_O(z_i)  & \emptyset & \emptyset & \emptyset &\{6\} & \{6,7\} & \{5,6,7,8\} & \{2,3,4,5,6\} & \{2,3,4,6\} \\
 M(z_i) & \emptyset & \emptyset & \emptyset & \{6\} & \{7\} & \{8,9,10\} & \{12\} & \emptyset
\end{array}$
\caption{Example of descriptions. $Z$ is the compressed string representing $Q$. We are looking for all matches of the pattern  $P$ with error threshold $k=2$ in $Z$.  The set of matches is $\{6,7,8,9,10,12\}$.}\label{fig:approx-description}
\end{center}
\end{figure}
Next we consider the complexity of computing the matches. To do this we first bound the size of the $M_I$ and $M_O$ sets. Since the length of any relevant suffix and relevant prefix is at most $m+k$, we have that $|M_O(z_i)|  \leq 2(m+k)<4m$, and therefore the total size of the $M_O$ sets is at most $O(nm)$. Each element in the sets $M_I(z_0), \ldots, M_I(z_n)$ corresponds to a unique match. Thus, the total size of the $M_I$ sets is at most $\occ$, where $\occ$ is the total number of matches. Since both sets are represented as sorted lists the total time to compute the matches for all compression elements is $O(nm + \occ)$.

\subsection{Computing Descriptions}

Next we show how to efficiently compute the descriptions. Let $1\leq \tau \leq n$ be a parameter. Initially, we compute a subset $C$ of the elements in $Z$ according to Lemma~\ref{lem:special} with parameter $\tau$. For each element $z_j \in C$ we store $l_j$, that is, the length of $\phrase(z_j)$. If $l_j > m+k$ we also store the index of the ancestor $x$ of $z_j$ of depth $m+k$. This information can easily be computed while constructing $C$ within the same time and space bounds, i.e., using $O(n\tau)$ time and $O(n/\tau)$ space.

Descriptions are computed from left-to-right as follows. Initially, set $l_0 = 0$, $u_0 = 0$, $\rpre(z_0) = \epsilon$, $\rsuf(z_0) = \epsilon$, $M_I(z_0) = \emptyset$, and $M_O(z_0) = \emptyset$. To compute the description of $z_i$, $1\leq i \leq n$, first
 follow the path $p$
 of references until we encounter an element $z_j \in C$. Using the information stored at $z_j$ we set $l_i := |p| + l_j$ and $u_i = u_{i-1} + l_{i-1}$. By Lemma~\ref{lem:special}(ii) the distance to $z_j$ is at most $2\tau$ and therefore $l_i$ and $u_i$ can be computed in $O(\tau)$ time given the description of $z_{i-1}$.

To compute $\rpre(z_i)$ we 
compute the label of the path from $z_0$ towards $z_i$ of length $\min(m+k, l_i)$. There are two cases to consider: 
If $l_i \leq m+k$ we simply compute the label of the path from $z_i$ to $z_0$ and let $\rpre(z_i)$ be the reverse of this string. Otherwise ($l_i > m+k$), we use the "shortcut" stored at $z_j$ to find the ancestor $z_h$ of distance $m+k$ to $z_0$. The reverse of the label of the path from $z_h$ to $z_0$ is then $\rpre(z_i)$. Hence, $\rpre(z_i)$ is computed in $O(m+k + \tau) = O(m + \tau)$ time.

The string $\rsuf(z_i)$ may be the divided over several phrases and we therefore recursively follow paths towards the root until we have computed the entire string. It is easy to see that the following algorithm correctly decodes the desired substring of length $\min(m+k, u_i)$ ending at position $u_i+l_i-1$.
\begin{enumerate}
  \item Initially, set $l := \min(m+k, u_i + l_i-1)$, $t:=i$, and $s := \epsilon$. 
  \item Compute the path $p$ of references from $z_t$ of length $r = \min(l, \depth(z_t))$ and set $s := s \cdot \lab(p)$.	
  \item If $r < l$ set $l := l-r$, $t := t - 1$, and repeat step $2$. 
  \item Return $\rsuf(z_i)$ as the reverse of $s$. 
\end{enumerate}
Since the length of $\rsuf(z_i)$ is at most $m+k$, the algorithm finds it in $O(m + k) = O(m)$ time.

The match sets $M_I$ and $M_O$ are computed as follows. Let $t(m,u,k)$ and $s(m,u,k)$ denote the time and space to compute $\Gamma(A,B)$ with error threshold $k$ for strings $A$ and $B$ of lengths $m$ and $u$, respectively. Since $|\rsuf(z_{i-1})\cdot \rpre(z_i)| \leq 2m+2k$ it follows that $M_O(z_i)$ can be computed in $t(m, 2m+2k,k)$ time and $s(m, 2m+2k,k)$ space. Since $M_I(z_i) = \Gamma(P, \phrase(z_i))$ we have that $j \in M_I(z_i)$ if and only if $j \in M_I(\reference(z_i))$ or $j =l_i$. By Proposition~\ref{prop:match} any match ending in $l_i$ must start within $[\max(1, l_i-m+1-k), \min(l_i, l_i-m+1+k)]$. Hence, there is a match ending in $l_i$ if and only if $l_i \in \Gamma(P, \rsuf'(z_i))$ where $\rsuf'(z_i)$ is the suffix of $\phrase(z_i)$ of length $\min(m+k, l_i)$. Note that $\rsuf'(z_i)$ is a suffix of $\rsuf(z_i)$ and we can therefore trivially compute it in $O(m+k)$ time. Thus, 
\begin{equation*}
M_I(z_i) = M_I(\reference(z_i)) \cup \{l_i \mid l_i \in  \Gamma(P, \rsuf'(z_i))\}\;.
\end{equation*}
Computing $\Gamma(P, \rsuf'(z_i))$ uses $t(m,m+k,k)$ time and $s(m, m+k, k)$ space. Subsequently, constructing $M_I(z_i)$ takes $O(|M_I(z_i)|)$ time and space. Recall that the elements in the $M_I$ sets correspond uniquely to matches in $Q$ and therefore the total size of the sets is $\occ$. Therefore, using dynamic perfect hashing~\cite{DKMMRT1994} on pointers to non-empty $M_I$ sets we can store these using $O(\occ)$ space in total.

\subsection{Analysis}
Finally, we can put the pieces together to obtain the final algorithm. The preprocessing uses $O(n\tau)$ expected time and $O(n/\tau)$ space. The total time to compute all descriptions and report occurrences is expected $O(n(\tau + m + t(m, 2m+2k,k)) + \occ)$. The description for $z_i$, except for $M_I(z_i)$, depends solely on the description of $z_{i-1}$. Hence, we can discard the description of $z_{i-1}$, except for $M_I(z_{i-1})$, after processing $z_i$ and reuse the space. It follows that the total space used is $O(n/\tau + m + s(m,2m+2k, k) + \occ)$. This completes the proof of Theorem~\ref{thm:approx}. Note that if we use $\Omega(n)$ space we can explicitly construct the dictionary. In this case hashing is not needed and the bounds also hold for the \zlw\ compression scheme.

\journal{
\section{Regular Expression Matching}\label{regular}

\subsection{Regular Expressions and Finite Automata}
First we briefly review the classical concepts used in the paper. For more details see, e.g., Aho et al.~\cite{ASU1986}. The set of \emph{regular expressions} over $\Sigma$ are defined recursively as follows: A character $\alpha \in \Sigma$ is a regular expression, and if $S$ and $T$ are regular expressions then so is the
  \emph{concatenation}, $(S)\cdot(T)$, the \emph{union}, $(S)|(T)$, and the \emph{star}, $(S)^*$. The \emph{language} $L(R)$ generated by $R$ is defined as follows: $L(\alpha) = \{\alpha\}$, $L(S \cdot T) = L(S)\cdot L(T)$, that is, any string formed by the concatenation of a string in $L(S)$ with a string in $L(T)$, $L(S)|L(T) = L(S) \cup L(T)$, and $L(S^*) = \bigcup_{i \geq 0} L(S)^i$, where $L(S)^0 = \{\epsilon\}$ and $L(S)^i = L(S)^{i-1} \cdot L(S)$, for $i > 0$.     
  
A \emph{finite automaton} is a tuple $A = (V, E, \Sigma, \theta, \Phi)$, where $V$ is a set of nodes called \emph{states}, $E$ is set of directed edges between states called \emph{transitions} each labeled by a character from $\Sigma \cup \{\epsilon\}$, $\theta \in V$ is a \emph{start state}, and $\Phi \subseteq V$ is a set of \emph{final states}. In short, $A$ is an edge-labeled directed graph with a special start node and a set of accepting nodes. 
$A$ is a \emph{deterministic finite automaton} (DFA) if $A$ does not contain any $\epsilon$-transitions, and all outgoing transitions of any state have different labels. Otherwise, $A$ is a \emph{non-deterministic automaton} (NFA). 

The \emph{label} of a path $p$ in $A$ is the concatenation of labels on the transitions in $p$. For a subset $S$ of states in  $A$ and character $\alpha \in \Sigma \cup \{\epsilon\}$, define the \emph{transition map}, $\delta(S, \alpha)$, as the set of states reachable from $S$ via a path labeled $\alpha$. Computing the set $\delta(S, \alpha)$ is called a \emph{state-set transition}. We extend $\delta$ to strings by defining $\delta(S, \alpha \cdot B) = \delta(\delta(S, \alpha), B)$, for any string $B$ and character $\alpha \in \Sigma$. We say that $A$ \emph{accepts} the string $B$ if $\delta(\{\theta\}, B) \cap \Phi \neq \emptyset$. Otherwise $A$ \emph{rejects} $Q$. As in the previous section, we say that $j \in [1, |B|]$ is a \emph{match} iff there is an $i \in [1, j]$ such that $A$ accepts $B[i,j]$. The set of all matches is denoted $\Delta(A, B)$.

Given a regular expression $R$, an NFA $A$ accepting precisely the strings in $L(R)$ can be obtained by several classic methods~\cite{MY1960, Glushkov1961, Thomp1968}. In particular, Thompson~\cite{Thomp1968} gave a simple well-known construction which we will refer to as a \emph{Thompson NFA} (TNFA). A TNFA $A$ for $R$ has at most $2m$ states, at most $4m$ transitions, and can be computed in $O(m)$ time. Hence, a state-set transition can be computed in $O(m)$ time using a breadth-first search of $A$ and therefore we can test acceptance of $Q$ in $O(um)$ time and $O(m)$ space. This solution is easily adapted to find all matches in the same complexity by adding the start state to each of the computed state-sets immediately before computing the next. Formally, $\deltab(S, \alpha \cdot B) = \deltab(\delta(S \cup \{\theta\}, \alpha), B)$, for any string $B$ and character $\alpha \in \Sigma$. 
A match then occurs at position $j$ if $\deltab(\{\theta\}, Q[1,j]) \cap \Phi \neq \emptyset$.

\subsection{Searching for Matches}

Let $A = (V, E, \Sigma, \theta, \Phi)$ be a TNFA with $m$ states. Given a compressed string $Z = z_1\ldots z_n$ representing a string $Q$ of length $u$ we show how to find $\Delta(A, Q)$ efficiently. As in the previous section let $l_i$ and $u_i$, $0\leq i \leq n$ be the length and start position of $\phrase(z_i)$. We process $Z$ from left-to-right and compute a description for $z_i$ consisting of the following information.
\begin{itemize}
  \item The integers $l_i$ and $u_i$.
  \item The state-set $S_{u_i} = \deltab(\{\theta\}, Q[1,u_i]+l_i-1)$.
  \item For each state $s$ of $A$ the compression element $\lastmatch(s, z_i) = x$, where $x$ is the ancestor of $z_i$ of maximum depth such that $\deltab(\{s\}, \phrase(x)) \cap \Phi \neq \emptyset$. If there is no ancestor that satisfies this, then $\lastmatch(s,z_i) = \bot$. 
\end{itemize}
\journal{\begin{figure}[t] 
  \centering \includegraphics[scale=.6]{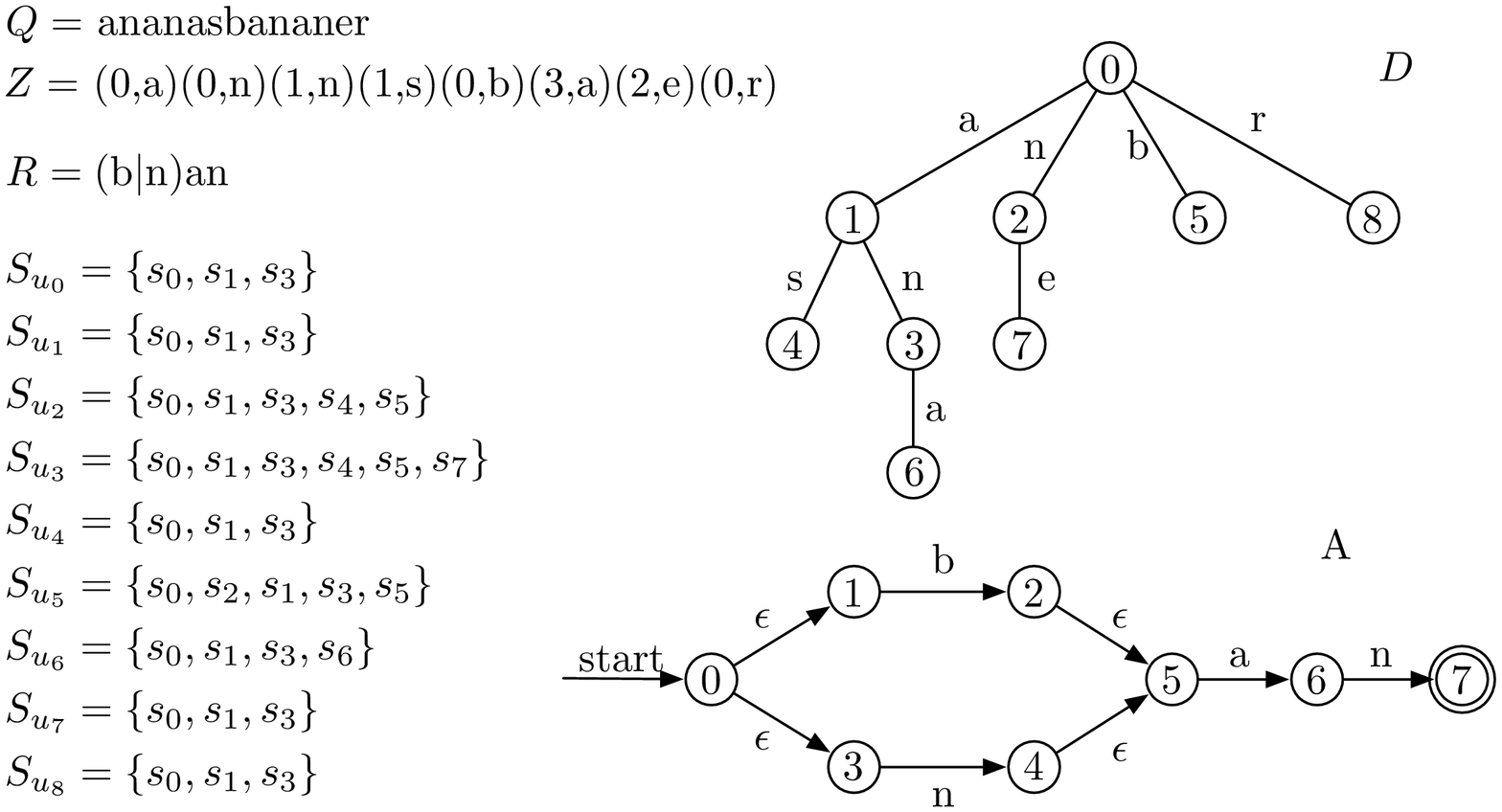}
  \caption{The compressed string $Z$ representing $Q$ and the corresponding dictionary trie $D$. The TNFA $A$ for the regular expression $R$ and the corresponding state-sets $S_{u_i}$ are given.  The lastmatch pointers are as follows: $\lastmatch(s_7,z_i)=\{z_0\}$ for $i=0,1,\ldots,8$, $\lastmatch(s_2,z_i)=\lastmatch(s_4,z_i)= \lastmatch(s_5,z_i)= \{z_3\}$ for $i=3,6$, and $\lastmatch(s_6,z_i)=  \{z_2\}$ for $i=2,7$. All other lastmatch pointers are $\bot$.
  Using the description we can find the matches: Since $s_2 \in S_{u_5}$ the element $z_3 \in M(s_2,z_6)$ represents the match $u_6+\depth(z_3)-1=9$. The other matches can be found similarly.}
  \label{fig:regex}
\end{figure}
An example description is shown in Fig.~\ref{fig:regex}.}{} The total size of the description for $z_i$ is $O(m)$ and therefore the space for all descriptions is $O(nm)$. In the next section we will show how to compute the descriptions. Assume for now that we have processed $z_0, \ldots, z_{i-1}$. We show how to find the matches within $[u_{i}, u_{i} + l_i - 1]$. Given a state $s$ define $M(s, z_i) = \{x_1, \ldots, x_k\}$, where $x_1 = \lastmatch(s, z_i)$, $x_j = \lastmatch(s, \parent(x_{j-1}))$, $1 < j \leq k$, and $\lastmatch(s, x_{k}) = \bot$, i.e., $x_1, \ldots, x_k$ is the sequence of ancestors of $z_i$ obtained by recursively following $\lastmatch$ pointers. By the definition of $\lastmatch$ and $M(s,z_i)$ it follows that $M(s,z_i)$ is the set of ancestors $x$ of $s$ such that $\deltab(s, x) \cap \Phi \neq \emptyset$. Hence, if $s \in S_{u_{i-1}}$ then each element $x \in M(s, z_i)$ represents a match, namely, $u_i + \depth(x)-1$. Each match may occur for each of the $|S_{u_{i-1}}|$ states and to avoid reporting duplicate matches we use a priority queue to merge the sets $M(s,z_i)$ for all $s \in S_{u_{i-1}}$, while generating these sets in parallel. A similar approach is used in~\cite{Navarro2003}. This takes $O(\log m)$ time per match. Since each match can be duplicated at most $|S_{u_{i-1}}| = O(m)$ times the total time for reporting matches is $O(\occ \cdot m\log m)$.

\subsection{Computing Descriptions}

Next we show how to compute descriptions efficiently. Let $1 \leq \tau \leq n$ be a parameter. Initially, compute a set $C$ of compression elements according to Lemma~\ref{lem:special} with parameter $\tau$. For each element $z_j \in C$ we store $l_j$ and the \emph{transition sets} $\deltab(s, \phrase(z_j))$ for each state $s$ in $A$. Each transition set uses $O(m)$ space and therefore the total space used for $z_j$ is $O(m^2)$. During the construction of $C$ we compute each of the transition sets by following the path of references to the nearest element $y\in C$ and computing state-set transitions from $y$ to $z_j$. By Lemma~\ref{lem:special}(ii) the distance to $y$ is at most $2\tau$ and therefore all of the $m$ transition sets can be computed in $O(\tau m^2)$ time.  Since, $|C| = O(n/\tau)$ the total preprocessing time is $O(n/\tau \cdot \tau m^2 ) = O(nm^2)$ and the total space is $O(n/\tau \cdot m^2)$.

The descriptions can now be computed as follows. The integers $l_i$ and $u_i$ can be computed as before in $O(\tau)$ time. All $\lastmatch$ pointers for all compression elements can easily be obtained while computing the transitions sets. Hence, we only show how to compute the state-set values. First, let $S_{u_0} := \{\theta\}$. To compute $S_{u_i}$ from $S_{u_{i-1}}$ we compute the path $p$ to $z_i$ from the nearest element $y \in C$. Let $p'$ be the path from $z_0$ to $y$. Since $\phrase(z_i) = \lab(p') \cdot \lab(p)$ we can compute $S_{u_i} = \deltab(S_{u_{i-1}}, \phrase(z_i))$ in two steps as follows. First compute the set 
\begin{equation}
\label{eq:union}
S' = \bigcup_{s \in S_{u_{i-1}}} \deltab(s, \phrase(y))\;.
\end{equation}
Since $y \in C$ we know the transition sets $\deltab(s, \phrase(y))$ and we can therefore compute the union in $O(m^2)$ time. Secondly, we compute $S_{u_i}$ as the set $\delta(S', \lab(p))$. Since the distance to $y$ is at most $\tau$ this step uses $O(\tau m)$ time. Hence, all the state-sets $S_{u_0}, \ldots, S_{u_n}$ can be computed in $O(nm(m + \tau))$ time.

\subsection{Analysis}
Combining it all, we have an algorithm using $O(nm(m + \tau) + \occ \cdot m \log m)$ time and $O(nm + nm^2/\tau)$ space. Note that since we are using $\Omega(n)$ space, hashing is not needed and the algorithm works for \zlw\ as well. In summary, this completes the proof of Theorem~\ref{thm:regularex}. 
\journal{
\subsection{Exploiting Word-level Parallelism}\label{sec:wordparallel}
If we use the word-parallelism inherent in the word-RAM model, the algorithm of Navarro~\cite{Navarro2003} uses $O(\ceil{m/w}(2^m + nm) + \occ\cdot m \log m)$ time and $O(\ceil{m/w}(2^m + nm))$ space, where $w$ is the number of bits in a word of memory and space is counted as the number of words used. The key idea in Navarro's algorithm is to compactly encode state-sets in bit strings stored in $O(\ceil{m/w})$ words. Using a DFA based on a Glushkov automaton~\cite{Glushkov1961} to  quickly compute state-set transitions, and bitwise OR and AND operations to compute unions and intersections among state-sets, it is possible to obtain the above result. The $O(\ceil{m/w}2^m)$ term in the above bounds is the time and space used to construct the DFA.

A similar idea can be used to improve Theorem~\ref{thm:regularex}. However, since our solution is based on Thompson's automaton we do not need to construct a DFA. More precisely, using the state-set encoding of TNFAs given in~\cite{Myers1992, BFC2005} a state-set transition can be computed in $O(\ceil{m/\log n})$ time after $O(n)$  time and space preprocessing. Since state-sets are encoded as bit strings each transition set uses $\ceil{m/\log n}$ space and the union in \eqref{eq:union} can be computed in $O(m\ceil{m/\log n})$ time using a bitwise OR operation. As $n \geq \sqrt{u}$ in \zla\  and \zlw, we have that $\log n \geq \frac{1}{2}\log u$ and therefore Theorem~\ref{thm:regularex} can be improved by roughly a factor $\log u$. Specifically, we get an algorithm using $O(n\ceil{m/\log u}(m + \tau)+ \occ\cdot m \log m)$ time and $O(nm\ceil{m/\log u}/\tau + nm)$ space.
}{Using simple word-parallel techniques and results from~\cite{Myers1992, BFC2005} it is straighforward to improve the bound by a factor roughly $\log u$. Due to lack of space we leave the details for the full version of this paper.}
}{}

\bibliographystyle{abbrv}
\bibliography{ref}

\end{document}